\documentclass[12pt,a4]{article}
\usepackage[dvips]{epsfig}
\renewcommand{\thefootnote}{\fnsymbol{footnote}}
\newcommand{\Li}[2]{{\mbox{Li}}_{#1}\left(#2\right)}
\def\gsim{\:\raisebox{-0.5ex}{$\stackrel{\textstyle>}{\sim}$}\:}
\begin{document}

\thispagestyle{empty}
\begin{flushright}
{MZ-TH/01-40} \\[5mm]
{hep-ph/0203125} \\[5mm]
{March 2002}           \\
\end{flushright}
\vspace*{3cm}    

\begin{center}
{\Large \bf
Hadronic corrections to the muon decay\footnote{
Talk presented by A.~D.
at the 16th International Workshop 
on High Energy Physics and Quantum Field Theory 
(Moscow, Russia, September 2001).
Supported 
by the DFG and by the Graduiertenkolleg ``Eichtheorien''
at the University of Mainz.}} \\
 
\vspace{10mm}
 
A.~I.~Davydychev$^{\; a,b,}$\footnote{
    Email: davyd@thep.physik.uni-mainz.de}~,
K.~Schilcher$^{\; a}$ and
H.~Spiesberger$^{\; a,}$\footnote{
    Email: hspiesb@thep.physik.uni-mainz.de} \\

\vspace{10mm}

${}^{a}$~Institut f\"ur Physik, Johannes-Gutenberg-Universit\"at,\\
Staudinger Weg 7, D-55099 Mainz, Germany \\[3mm]

${}^{b}$~Institute for Nuclear Physics, Moscow State University,\\
119992 Moscow, Russia

\end{center}
 
\vspace{15mm}
 
\begin{abstract}
\noindent
We consider the ${\cal O}(\alpha^2)$ hadronic corrections to
the energy spectrum of the decay electron in muon decay. We find that
the correction can be described, within good approximation, by a linear
function in the electron energy. Explicit expressions for the form
factors needed in an approach based on dispersion integrals are given.
\end{abstract}

\renewcommand{\thefootnote}{\arabic{footnote}}
\setcounter{footnote}{0}
\newpage 



At present, the best value of the muon
life time is $\tau_{\mu}=(2.19703\pm0.00004) \, \mu s$ \cite{PDG}.
In fact, experimental data have reached such a precision that quantum
corrections have to be taken into account when comparing this 
experimental result with theoretical predictions.
To match this accuracy from the theory
side, two-loop radiative corrections to the muon decay in the full
electroweak Standard Model are needed. 
To perform the required calculations is a formidable, but not
impossible, task. A step in this direction is the calculation of the
purely electromagnetic corrections to 
the total decay rate at order $\alpha^{2}$ in the Fermi
theory, which has been performed by van Ritbergen and Stuart
\cite{Rstuart1,Rstuart2} (see also \cite{steinhauser}). Besides that,
also the ${\cal O}(N_f \alpha^2)$ corrections in the Standard Model have
been considered in Refs.\ \cite{Malde,Freitas}.
The decay spectrum is known to order $\alpha$ since long; however, the
corresponding corrections for 
polarized muon decay were calculated only recently
in~\cite{polarized}. The leading logarithmic QED corrections of order 
${\cal O}\left(\alpha^2\ln^2(m_{\mu}/m_e)\right)$ 
to the muon decay spectrum were considered in Ref.~\cite{ACzG}. 

The spectrum calculation is
different from the one for the life time since the
Kinoshita-Lee-Nauenberg theorem \cite{KLN} is not in effect.
Consequently, powers of the large logarithm $\ln(m_{\mu}/m_{e})$ do not
cancel in the calculation of the electromagnetic corrections to the
spectrum. This becomes obvious when fitting the spectrum corrected to
order ${\cal O}(\alpha)$ to the Michel spectrum: the resulting effective
Michel parameter differs by about $6\,\%$ from its lowest-order value
\cite{drho}, a correction which is more than 10 times larger than the
corresponding correction to the muon life time. At order 
${\cal O}(\alpha^{2})$ the radiative corrections may be expected to be of 
the order of several per-mille 
and will become important for future high-precision experiments. 

Given this perspective we review the calculation of the
hadronic contribution to the energy spectrum of the final-state electron
in muon decay. This contribution is not expected to be logarithmically
enhanced, but nonetheless is required for an eventual complete
second-order calculation. Further details of the calculation are given 
in Ref.~\cite{DSS}.



We consider the decay of a muon in its rest system,  
\begin{equation}
\mu^-(p) \rightarrow e^-(p^{\prime}) + \nu_{\mu}(q_1) + \bar{\nu}_e(q_2)
\, , 
\label{mdecay}
\end{equation}
and define momenta as shown in (\ref{mdecay}).  
It is convenient to introduce the dimensionless variable
$x=2E_e/m$ to denote the ratio of the energy of the decay electron 
$E_e$ with
respect to the muon mass $m_{\mu}\equiv m$. Neglecting the electron mass,
$p^{\prime 2} = 0$, the kinematically allowed range is
$0 \le x \le 1$, and the momentum transferred
from the charged particles to the neutrino pair, $q = q_1 + q_2 = p -
p^{\prime}$ is determined by 
$q^2 = (1 - x) m^2$.

The matrix element ${\cal M}$ for (\ref{mdecay}) in the Fermi theory can
be calculated most conveniently after a Fierz rearrangement factorizing
the amplitude into a current $J_{\mu}$ which describes the $\mu e$
transition and a current for the $\nu_{\mu} \nu_e$ interaction. After
squaring and summing (averaging) over spins, one can write 
$|{\cal M}|^2$ as a product of two corresponding tensors. 
The one pertaining
to the neutrino interaction can be integrated over the unobserved
momenta of the neutrinos independently, leading to
\begin{equation}
N_{\mu \nu} = q_{\mu} q_{\nu} - g_{\mu \nu} q^2 \, .
\label{nmunu}
\end{equation}
This tensor will be contracted with 
$C_{\mu \nu} = J^{\ast}_{\mu} J_{\nu}$, 
with 
$J_{\mu} = \bar{u}_e(p^{\prime}) \Lambda_{\mu}(q) u_{\mu}(p)$,
where $\Lambda_{\mu}(q)$ is the effective vertex of the four-fermion
interaction.  At the lowest order, $\Lambda_{\mu}(q)$ is identified with
\begin{equation}
\Lambda_{\mu}^0 = \frac{G_F}{\sqrt{2}} \gamma_{\mu} 
\left(1 - \gamma_5\right) \, ,
\end{equation}
where $G_F = (1.16637 \pm 0.00001) \times 10^{-5}$ GeV$^{-2}$ is the
Fermi coupling constant.

\begin{figure}[tb]
\unitlength 1mm
\vspace{-1cm}
\begin{picture}(120,53)
\put(5,0){\epsfig{file=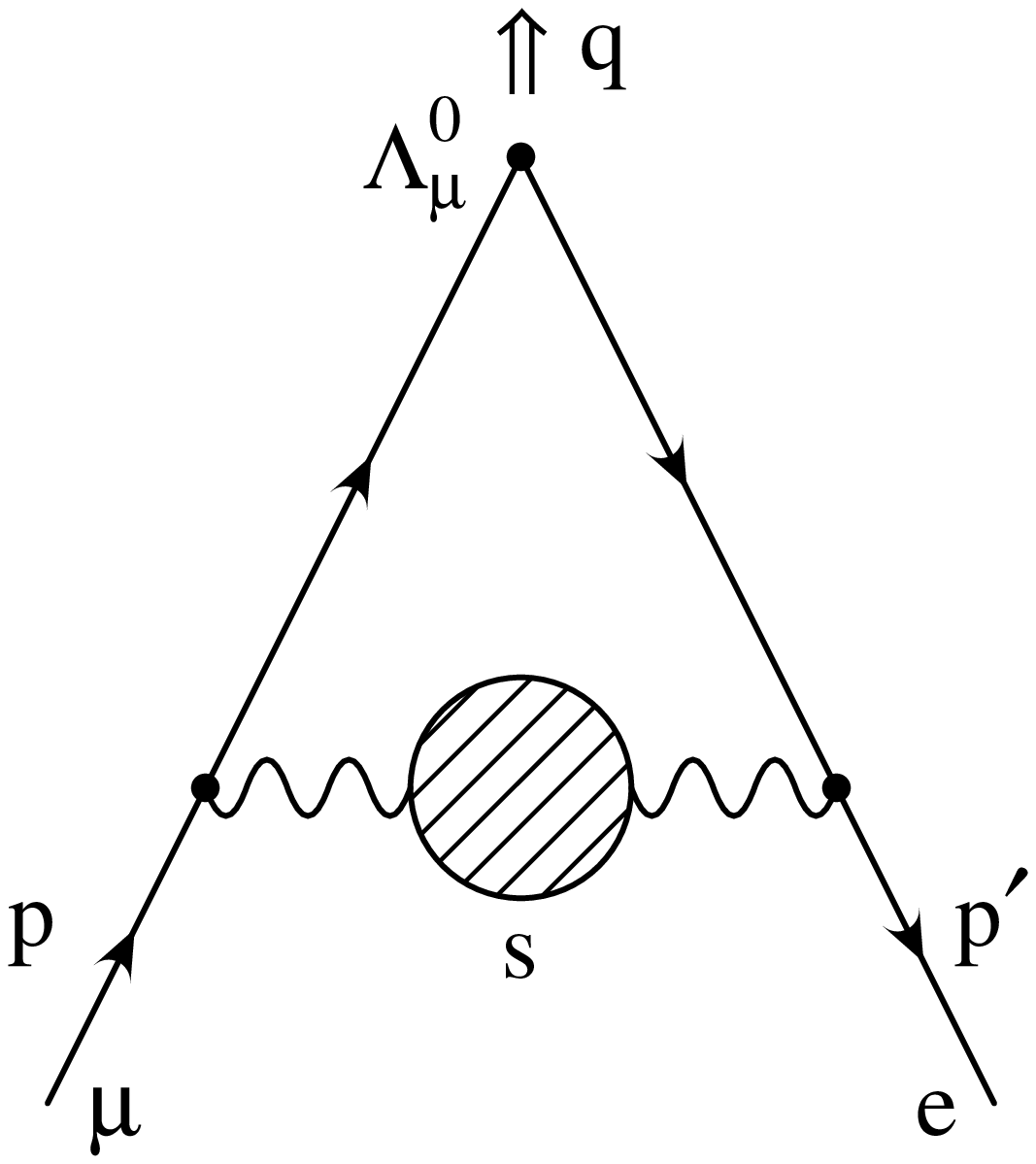,,width=4.97cm}}
\put(55,0){\epsfig{file=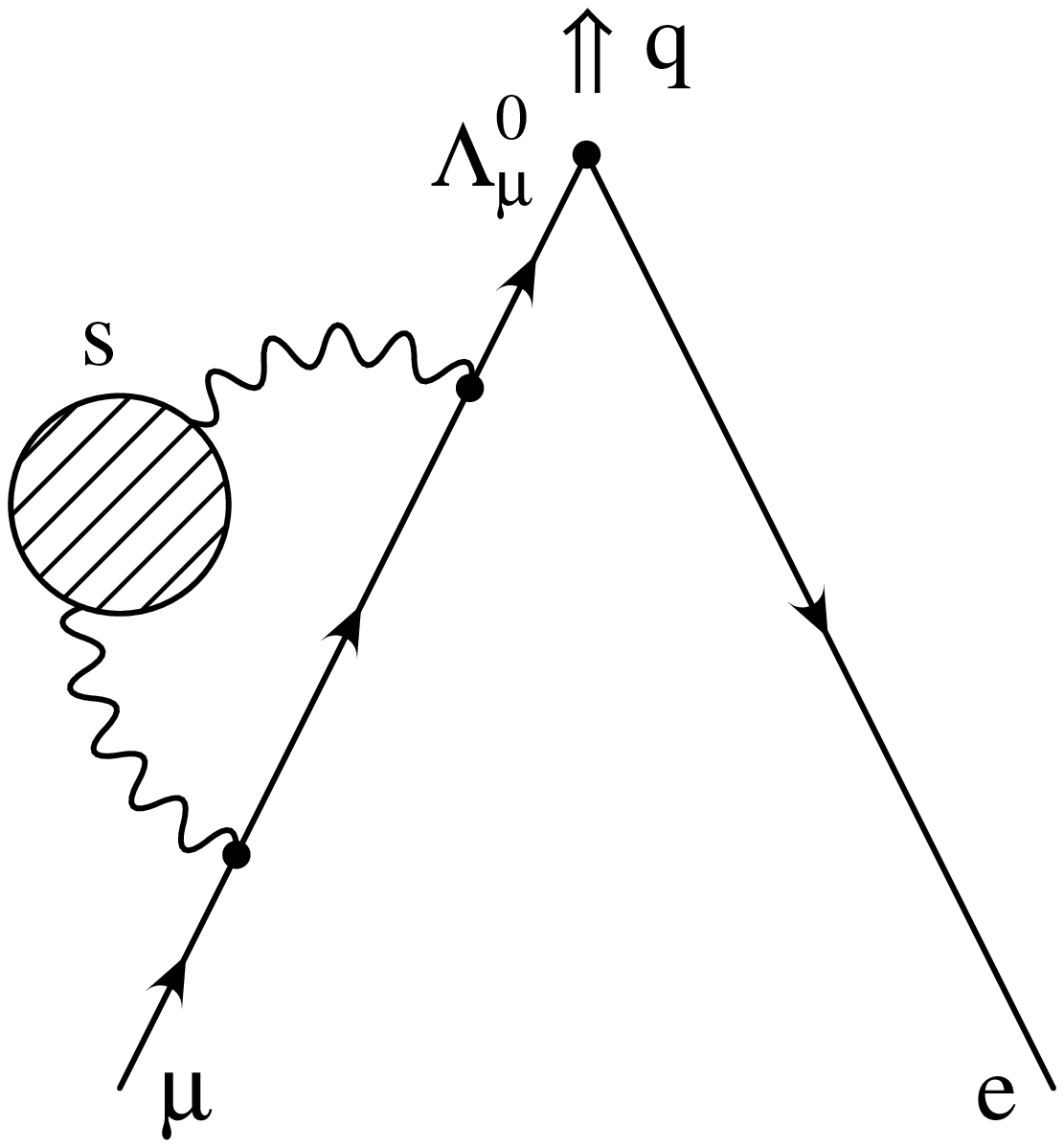,,width=4.9cm}}
\put(110,0){\epsfig{file=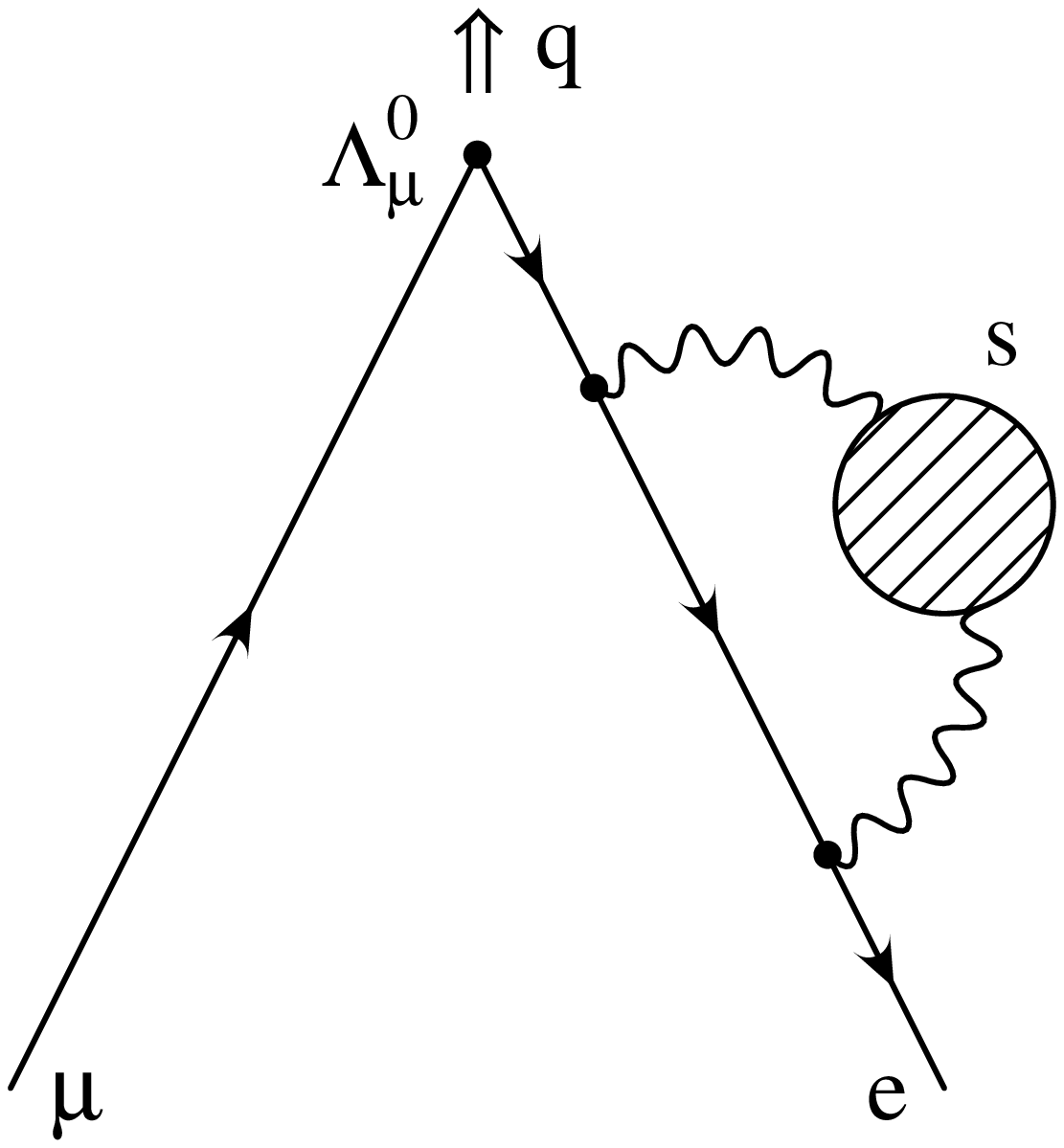,,width=4.9cm}}
\put(27,-2){a}
\put(82,-2){b}
\put(131,-2){c}
\end{picture}
\caption{\it Feynman diagrams describing a self energy insertion in the
  photonic one-loop corrections to the $\mu e$ vertex.}
\label{fig_fd}
\end{figure}

Hadronic contributions to the radiative corrections to the current
$J_{\mu}$ at order ${\cal O}(\alpha^2)$ are described by the Feynman
diagrams shown in Fig.\ \ref{fig_fd}. The hadronic vacuum polarization
\begin{equation}
\Pi^{\rm had}_{\mu \nu} (k^2) = 
\frac{-{\rm i} \Pi^{\rm had}(k^2)}{k^2 + {\rm i}0}
\left( g_{\mu \nu} - \frac{k_{\mu} k_{\nu}}{k^2} \right)
\end{equation}
is inserted in a one-loop vertex correction. The vacuum polarization can
be related to the measured cross section for $e^+e^- \rightarrow
hadrons$ with the help of a dispersion relation
\begin{equation}
\Pi^{\rm had}(k^2) = \frac{\alpha}{3\pi} \int_{s_{\rm thr}}^{\infty}
\frac{{\rm d}s}{s} R(s) \frac{k^2}{k^2 - s  + {\rm i}0} \, ,
\label{dispersion-int}
\qquad 
R(s) = \frac{\sigma(s; e^+e^- \rightarrow hadrons)}%
            {\sigma(s; e^+e^- \rightarrow \mu^+ \mu^-)} ,
\end{equation}
where the integration starts at the two-pion threshold, $s_{\rm thr} =
4 m_{\pi}^2$.  Therefore, the calculation corresponds to a one-loop
vertex with a photon of mass $\sqrt{s}$, i.e.\ using a
propagator
\begin{equation}
\frac{-{\rm i}}{k^2 - s + {\rm i}0}
\left( g_{\mu \nu} - \frac{k_{\mu}k_{\nu}}{k^2} \right) \, .
\end{equation}
The result can be written in the form
\begin{equation}
\Lambda_{\mu}(q) = \frac{\alpha}{3\pi} \int_{s_{\rm thr}}^{\infty}
\frac{{\rm d}s}{s} R(s) \widetilde{\Lambda}_{\mu}(s; q, m^2) 
\end{equation}
where the vector function $\widetilde{\Lambda}_{\mu}$ can be decomposed
into Lorentz-covariants as
\begin{equation}
\widetilde{\Lambda}_{\mu}(s; q, m^2) = 
\gamma_{\mu} \omega_L \left[ 1 + \tilde{f}(s; q^2, m^2)\right] + 
\frac{p_{\mu}+p^{\prime}_{\mu}}{m} \omega_R \tilde{g}_+(s; q^2, m^2) +
\frac{q_{\mu}}{m} \omega_R \tilde{g}_-(s; q^2, m^2) \, ,
\label{form-factor}
\end{equation}
with $\omega_{R,L} = (1 \pm \gamma_5)/2$. The calculation is
straightforward and corresponds to that of the one-loop vertex
correction in QED, with the difference that ({\it i}) the exchanged
``photon'' is massive with mass $\sqrt{s}$, ({\it ii}) the coupling is
purely left-handed, and ({\it iii}) the two fermion lines have different
masses. In fact, except for small $q^2$ the electron mass can safely be
neglected.  After contraction with the
neutrino tensor $N_{\mu \nu}$, Eq.\ (\ref{nmunu}), only the form factors
$\tilde{f}(s; q^2, m^2)$ and $\tilde{g}_+(s; q^2, m^2)$ will remain in
the final result.

Explicit results for the relevant form factors
$\tilde{f}$ and
$\tilde{g}_{+}$ are~\cite{DSS}
\begin{eqnarray}
\label{f}
\tilde{f}(s; q^2, m^2) & = &
- \frac{\alpha}{4\pi} \Biggl\{ 
2(m^2-q^2-s) \left[ 1+\frac{q^2 s}{(m^2-q^2)^2} \right] C_0  
\nonumber \\
&& + \left[ \frac{2 q^2 s}{(m^2-q^2)^2}
           -\frac{2 q^2}{(m^2-q^2)}+1 \right] B_0(q^2;m^2,0)
\nonumber \\
&& + \left[ \frac{2 q^2}{m^2-q^2}
           -\frac{s(m^2+q^2)}{(m^2-q^2)^2}\right] B_0(m^2;m^2,s)
+\left( \frac{1}{s}-\frac{1}{m^2-q^2} \right) A(s)
+ 2 \Biggl\}
\nonumber \\
&& + \tilde{f}_{\rm SE}(s; q^2, m^2) \; ,
\\[1ex]
\label{g+}
\tilde{g}_{+}(s; q^2, m^2) &=&
- \frac{\alpha}{4\pi} \frac{m^2}{q^2}
\Biggl\{ \frac{2q^2 s}{m^2-q^2}
\left[ \frac{3q^2 s}{(m^2-q^2)^2} +2 \right] C_0
\nonumber \\
&& 
-\frac{q^2}{m^2-q^2}
    \left[ \frac{6 q^2 s}{(m^2-q^2)^2} + 1 \right] B_0(q^2;m^2,0) 
- \frac{q^2}{m^2-q^2}
\nonumber \\
&& + \frac{q^2}{m^2-q^2}
    \left[ \frac{6 m^2 s}{(m^2-q^2)^2} 
          -\frac{s(4m^2-q^2)}{m^2(m^2-q^2)} +2 \right] B_0(m^2;m^2,s)
\nonumber \\
&& + \frac{q^2}{m^2-q^2}
\left[ \frac{3}{m^2-q^2} - \frac{1}{m^2} \right] A(s)
+ \left[ \frac{1}{m^2-q^2} - \frac{1}{m^2} \right] A(m^2)
\Biggl\} \; ,
\end{eqnarray}
where $C_0 = C_0(m^2,0,q^2;m^2,s,0)$ is the three-point integral (cf.\ 
Fig.\ \ref{fig_fd}a), whereas $B_0$ and $A$
denote the tadpole and two-point integrals \cite{tHV'79,PV},
respectively (see Eqs.~(\ref{A})--(\ref{B2}) below).
The complete set of results, including that for  
$\tilde{g}_{-}$, is presented in~\cite{DSS}.

Self-energy diagrams (Fig.~\ref{fig_fd}b,c) contribute to the
coefficient of $\gamma_{\mu}$ only. The result is 
\begin{eqnarray}
\tilde{f}_{\rm SE}(s; q^2, m^2) & = & 
\frac{\alpha}{8\pi} \Biggl\{
2(s + 2m^2) 
\left.\frac{\partial B_0(p^2;m^2,s)}{\partial p^2}\right|_{p^2=m^2}
- \frac{s}{m^2} B_0(m^2;m^2,s)
\nonumber \\
&& - \frac{s+m^2}{s m^2} A(s) + \frac{1}{m^2} A(m^2) + 1
+ \frac{3}{2} \Biggl\} ,
\label{fse}
\end{eqnarray}
where the last term, ${\textstyle{3\over2}}$, comes from the self energy
on the massless (electron) leg. Using recurrence relations~\cite{ibp}
(see also Appendix~A of~\cite{BDS}), the derivative in (\ref{fse}) can
be represented as
\[
\frac{\partial B_0(p^2;m^2,s)}{\partial p^2}\Biggl|_{p^2=m^2}
= \frac{1}{m^2 (s \!-\! 4m^2)}
\Biggl[\! m^2 -(s-3m^2) B_0(m^2;m^2,s) 
+ A(m^2) - \left( \! 1 \!-\! \frac{2m^2}{s} \! \right) \! A(s)\! \Biggl]\! .
\]
The required tadpole and two-point integrals are~\cite{tHV'79,PV}
\begin{eqnarray}
A(m^2) & = & m^2 \left[ - \Delta  - 1 
+ \ln \frac{m^2}{\mu_{\rm DR}^2} \right],
\label{A}
\\
B_0(m^2;m^2,s) & = & \Delta + 2 - \ln \frac{m^2}{\mu_{\rm DR}^2} 
- \frac{s}{2m^2} \ln \frac{s}{m^2}
+ \frac{s}{2m^2} \beta
\ln\left(\frac{1+\beta}{1-\beta}\right),
\label{B1}
\\
B_0(q^2;m^2,0) & = & \Delta + 2 -
\ln \frac{m^2}{\mu_{\rm DR}^2}
+ \frac{m^2-q^2}{q^2} \ln\left(\frac{m^2-q^2}{m^2}\right) ,
\label{B2}
\end{eqnarray}
where 
$\beta \equiv \sqrt{1-4m^2/s}$, 
and $\mu_{\rm DR}$ is the scale parameter of dimensional regularization.
In Eqs.~(\ref{A})--(\ref{B2}), terms containing $\Delta = 1/\varepsilon
- \ln\pi - \gamma_{\rm E}$ represent the ultraviolet singularities which
cancel in the final results (\ref{f})--(\ref{g+}).

Finally, we need the three-point scalar function $C_0$ \cite{tHV'79} for
positive values of $q^2$. For arbitrary values of $m_{\mu}$ and
$m_e$, it can be presented in the following symmetric form:
\begin{eqnarray}
\label{C0gen}
C_0 &=& \frac{1}{\sqrt{\lambda(m_e^2,{m_{\mu}}^2,q^2)}}
      \Biggl\{ -\Li{2}{1-\frac{c m_e}{m_{\mu}}}
               -\Li{2}{1-\frac{c m_{\mu}}{m_e}}
\nonumber \\ &&
          +\Li{2}{1-\frac{c b_{\mu}}{b_e}}
          +\Li{2}{1-\frac{c b_e}{b_{\mu}}}
          +\Li{2}{1-\frac{c}{b_e b_{\mu}}}
          +\Li{2}{1-c b_e b_{\mu}}
\nonumber \\ &&
          +\ln^2{b_e}+\ln^2{b_{\mu}}+\frac{1}{2}\ln^2{c}
          -\frac{1}{2}\ln^2{\frac{m_e}{m_{\mu}}}
          -\ln{c}\ln{\frac{m_e m_{\mu}}{s}} \Biggl\} \; ,  
\end{eqnarray}
where
\begin{eqnarray}
&& b_e \equiv \sqrt{\frac{1-\beta_e}{1+\beta_e}}, \quad
b_{\mu} \equiv \sqrt{\frac{1-\beta_{\mu}}{1+\beta_{\mu}}}, \quad
\beta_e\equiv \sqrt{1-\frac{4 m_e^2}{s}}, \quad
\beta_{\mu} \equiv \sqrt{1-\frac{4 {m_{\mu}}^2}{s}},
\\
&& c \equiv 
\sqrt{
\frac{m_e^2+{m_{\mu}}^2-q^2-\sqrt{\lambda(m_e^2,{m_{\mu}}^2,q^2)}}
{m_e^2+{m_{\mu}}^2-q^2+\sqrt{\lambda(m_e^2,{m_{\mu}}^2,q^2)}}}
\end{eqnarray}
and
$\lambda(x,y,z) \equiv x^2+y^2+z^2-2xy-2yz-2zx$ is the standard
notation for the K{\"a}llen function.
This expression is regular in the limit $m_e\to~0$. 
A compact result for this case is given in Eq.~(34)
of~\cite{DSS}. 


Performing the dispersion integral one obtains the form factors
\begin{equation}
f(q^2, m^2) = \frac{\alpha}{3\pi} \int_{s_{\rm thr}}^{\infty}
\frac{{\rm d}s}{s} R(s) \tilde{f}(s; q^2, m^2) \, ,
\quad
g_{+}(q^2, m^2) = \frac{\alpha}{3\pi} \int_{s_{\rm thr}}^{\infty}
\frac{{\rm d}s}{s} R(s) \tilde{g}_+(s; q^2, m^2) \, .
\label{formfactors}
\end{equation}
Since we are interested in the ${\cal O}(\alpha^2)$ correction, it is
sufficient to keep only terms of first order in $f$ and $g_+$ in the
decay spectrum which then can be written in the form
\begin{equation}
\frac{1}{\Gamma_0} \frac{{\rm d} \Gamma}{{\rm d} x} =  
2 x^2 \left[ (3 - 2x) \left(1 + 2f(x)\right) + x g_+(x) \right] = 
\left.\frac{1}{\Gamma_0} \frac{{\rm d} \Gamma}{{\rm d} x}\right|_{\rm Born} 
(1 + r(x)) \, ,
\label{spectrum}
\end{equation}
with  
\begin{equation}
\Gamma_0 = \frac{G_F^2 m^5}{192 \pi^3}\, ,
\qquad 
\left. \frac{{\rm d} \Gamma}{{\rm d} x}\right|_{\rm Born} = 
2 x^2 (3 - 2x) \Gamma_0\, , \qquad
r(x) = 2f(x) + \frac{x}{3 - 2x} g_+(x) \, ,
\label{rx}
\end{equation}
where $f(x) \equiv f\left((1-x)m^2,m^2\right)$ 
and $g_+(x) \equiv g_+\left((1-x)m^2,m^2\right)$. 




For our purpose, the function $R(s)$ describing the hadronic
cross section of $e^+e^-$ annihilation can be modeled by a combination
of experimental data and analytical results from perturbative QCD. Since
we are going to calculate a small correction, it is not necessary to
invoke the most sophisticated treatment as needed, for example, when
calculating the hadronic contribution to the fine structure constant
$\alpha(m_Z)$. At low $s < 2.5$ GeV$^2$ we use experimental data from
ALEPH parametrized in \cite{davier} or provided directly by ALEPH
\cite{aleph-tau} from a measurement of the isovector $\tau$ spectral
function. These data are complemented by the resonance contributions
from the isospin-0 light mesons $\omega$ and $\phi$. Above $s = 2.5$
GeV$^2$ we use the QCD prediction for $R(s)$ due to light quarks at
order ${\cal O}(\alpha_{\rm s})$.

Since the data in the $c\bar{c}$-channel published by various groups are
in a large part of the energy range inconsistent, we apply in this case
the QCD-based approach of analytic continuation by duality \cite{acd}.
The data region can be chosen to extend only over the sub-threshold
resonances, i.e.\ one can calculate the contribution coming from the
$c\bar{c}$-channel by a combination of data describing the $J/\Psi(1S)$
and $J/\Psi(2S)$ resonances and the prediction of perturbative QCD. We
checked that the results obtained this way are consistent with those of
the standard approach using the new BES data \cite{besii}. 



\begin{figure}[tb]
\unitlength 1mm
\vspace{-1cm}
\begin{picture}(120,110)
\put(30,0){\epsfig{file=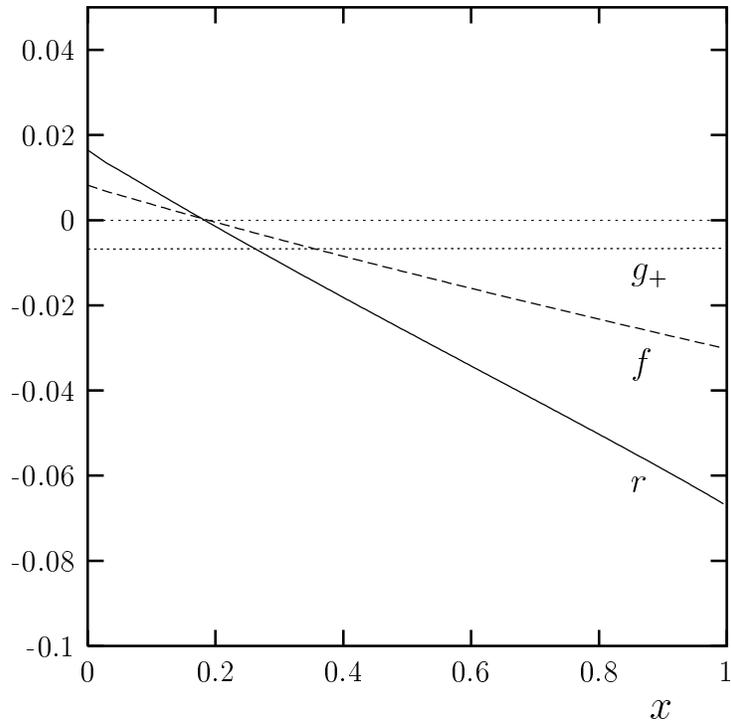,,width=10cm}}
\end{picture}
\caption{\it Results for the form factors $f$, $g_+$ and $r$ defined in
  (\ref{formfactors}), (\ref{rx}).} 
\label{fig_fgrx}
\end{figure}

The correction to the total decay rate, 
$\Delta \Gamma = \int_0^1 {\rm d}x \left({\rm d}\Gamma/{\rm d} x\right)$, 
was calculated before in \cite{Rstuart1}.  Our result,
\begin{equation}
\Delta \Gamma_{\rm had} \simeq - 0.0421
\left(\alpha/\pi\right)^2  \Gamma_0 \; ,
\end{equation}
agrees perfectly with the corresponding number $-0.042$ given in
\cite{Rstuart1}.  The resulting corrections to the spectrum
(\ref{spectrum}) are shown in Fig.\ \ref{fig_fgrx}. At small $x$, the
corrections are positive, but the correction to the total decay width is
dominated by the negative values at $x \gsim 0.18$. The dependence of
the form factors on $x$ is to a very good approximation linear:
\begin{eqnarray}
f(x) & \simeq & \left(0.0071 - 0.0378 x \right) 
\left(\alpha/\pi\right)^2 \Gamma_0 
\, , \nonumber\\
g_+(x) & \simeq & - 0.0067 \left(\alpha/\pi\right)^2 \Gamma_0  
 \, , \\
r(x) & \simeq & \left(0.0148 - 0.0813 x \right) 
\left(\alpha/\pi\right)^2 \Gamma_0 
\, . \nonumber
\end{eqnarray}
For $g_+$, the coefficient of the term linear in $x$ is very small and
is therefore omitted.  Note that this behaviour cannot be described by a
simple redefinition of the Michel parameter. Since $G_F$ is a free
parameter in the Fermi theory, the correction to the total decay width
is not observable; it can be absorbed by a suitable redefinition of the
Fermi constant.  However, the modification of the spectrum is, in
principle, observable.

\begin{table}[bt]
\begin{center}
\begin{tabular}{rlrr}
\hline\hline
\multicolumn{4}{c}{ Contributions to $\Delta \Gamma_{\rm had}$} \\
\hline
1 & $0 < s < 0.2$ GeV$^2$                   & $- 0.00129$ & 3.1 \% \\
2 & $\omega$                                & $- 0.00223$ & 5.3 \% \\
3 & $\phi$                                  & $- 0.00264$ & 6.3 \% \\
4 & $0.2 < s < 2.5$ GeV$^2$                 & $- 0.02804$ & 66.6 \% \\
5 & $s > 2.5$ GeV$^2$                       & $- 0.00564$ & 13.4 \% \\
6 & $J/\Psi(1S)$                            & $- 0.00066$ & 1.6 \% \\
7 & $J/\Psi(2S)$                            & $- 0.00017$ & 0.4 \% \\
8 & charm, $s > 4 m_c^2$                    & $- 0.00138$ & 3.3 \% \\
9 & bottom, $s > 4 m_b^2$                   & $- 0.00003$ & 0.1 \% \\
\hline
  & Sum                                     & $- 0.04207$ & 100 \% \\
\hline\hline
\end{tabular}
\caption{Contributions to the corrections of the total decay rate}
\label{tab_Gtot}
\end{center}
\end{table}

The correction to the total decay rate can be split up into the various
contributions to the hadronic vacuum polarization, as shown in Table
\ref{tab_Gtot}. The form factor $f$ of the $\gamma_{\mu}$ term
contributes $-0.0387$, whereas the correction due to $g_+$, $-0.0033$,
is smaller by one order of magnitude. The total correction (the
contributions due to $f$ and $g_+$) is saturated to 81\,\% (80.8\,\% and
89\,\%, respectively) by the contributions from small $s$ below 2.5
GeV$^2$.  Only 5\,\% of $\Delta \Gamma_{\rm had}$ is due to charmed
states, and the bottom sector is completely negligible.

The numerical results given here take into account the ${\cal
  O}(\alpha_{\rm s})$ QCD corrections in $R(s)$. Using the leading-order
expression for $R(s)$, i.e.\ without the correction factor $(1 +
\alpha_{\rm s}/\pi)$ in the light-quark contribution at large $s$ and
the charm-quark contribution, the final result would be $-0.04149$,
i.e.\ changed by 1.4\,\%. We conclude that a more refined treatment
which would include higher orders of perturbative QCD and mass-dependent
corrections in the heavy-quark sector is not required for our purpose.

The evaluation of the dispersion integrals has been performed using
standard numerical integration routines up to a value $s_{\rm max}$ of
several hundred GeV$^2$. The contribution above this value was obtained
with the help of {\tt Maple} using the asymptotic expansion of the
form factors for large $s$ (see in the Appendix of Ref.~\cite{DSS}). 
For intermediate values of $s$, good consistency of both
procedures has been verified.   


The same set of formulae can be used to calculate the
contributions from a $\mu^+ \mu^-$ or a $\tau^+ \tau^-$ loop insertion.
We find that the tau loop gives a very small contribution, about
$1.5\,\%$ of the one from the muon loop, in agreement with Ref.\ 
\cite{Rstuart1}. Therefore we give only results for the muon loop, where
one has to insert in Eq.\ (\ref{formfactors})
\begin{equation}
R(s) \rightarrow \left(1 + \frac{2 m^2}{s} \right) 
                 \sqrt{1 - \frac{4 m^2}{s}}
\quad {\rm and} \quad 
s_{\rm thr} \rightarrow 4 m^2\, .
\end{equation}
With this input we obtain 
\begin{equation}
\Delta \Gamma_{\rm muon} \simeq - 0.0364 \left(\alpha/\pi\right)^2
\Gamma_0 \; , 
\end{equation}
which perfectly agrees with the exact result given in 
Ref.~\cite{Rstuart1}.  The results of a linear fit of the form factors for
the muon-loop insertion are:
\begin{eqnarray}
f_{\rm muon}(x) & \simeq & \left(0.0130 - 0.0414 x \right) 
\left(\alpha/\pi\right)^2 \Gamma_0 \, , 
\nonumber\\
g_{+,{\rm muon}}(x) & \simeq & \left( - 0.0090 + 0.0005 x \right) 
\left(\alpha/\pi\right)^2 \Gamma_0 \, , \\
r_{\rm muon}(x) & \simeq & \left(0.0267 - 0.0898 x \right) 
\left(\alpha/\pi\right)^2 \Gamma_0 \, . \nonumber
\end{eqnarray}

When analysing the similar contribution with an $e^{+}e^{-}$ loop
insertion (which represents one of the pure QED contributions), one
should carefully treat logarithmic singularities occurring in the limit
$m_e\to0$.  For $m_e\neq0$, the general result (\ref{C0gen}) can be
used.   




One can see that the energy spectrum in the decay of
an unpolarized muon is corrected by a smooth function in $x$ due to
hadronic contributions at order ${\cal O}(\alpha^2)$. At both ends of
the spectrum no particularly outstanding enhancement or suppression is
observed. 

The calculations described in this paper constitute only the most
straightforward part of a full calculation which would be necessary
before the expected future high-precision data can be confronted with
theoretical predictions. This will not only be necessary for a
meaningful test of the electroweak Standard Model, but also when
searching for physics beyond the Standard Model \cite{kuno}. 


\end{document}